\documentclass[journal]{IEEEtran}
\usepackage{cite}
\usepackage{amsmath,amssymb,amsfonts,dsfont, mathrsfs}
\usepackage[noend]{algpseudocode}
\usepackage{algorithmicx,algorithm}

\usepackage{graphicx}
\usepackage{subfigure}
\usepackage{textcomp}
\usepackage{xcolor}
\usepackage{booktabs}
\usepackage{breqn}
\usepackage{subeqnarray}
\usepackage{cases}
\usepackage{setspace}
\usepackage{amsthm}
\usepackage{bbm}
\usepackage{bm}
\usepackage{lipsum}
\usepackage{enumitem}
\usepackage{makecell}
\usepackage{multirow}
\usepackage{pifont}
\usepackage{epstopdf}
\usepackage{hyperref}

\allowdisplaybreaks[4]

\theoremstyle{remark}

\newtheorem{rem}{Remark}

\def\BibTeX{{\rm B\kern-.05em{\sc i\kern-.025em b}\kern-.08em
    T\kern-.1667em\lower.7ex\hbox{E}\kern-.125emX}}

\begin{document}

\title{Sampling-Free Diffusion Transformers for Low-Complexity MIMO Channel Estimation\\
}
\author{Zhixiong~Chen,~\IEEEmembership{Member,~IEEE},
Hyundong~Shin,~\IEEEmembership{Fellow,~IEEE},\\
and Arumugam~Nallanathan,~\IEEEmembership{Fellow,~IEEE}
\thanks{Zhixiong Chen is with the School of Electronic Engineering and Computer Science, Queen Mary University of London, E1 4NS London, U.K. (email: zhixiong.chen@qmul.ac.uk).}
\thanks{Hyundong Shin is with the Department of Electronics and Information Convergence Engineering, Kyung Hee University, Yongin-si, Gyeonggido 17104, Republic of Korea (e-mail: hshin@khu.ac.kr).}
\thanks{Arumugam Nallanathan is with the School of Electronic Engineering and Computer Science, Queen Mary University of London, E1 4NS London, U.K., and also with the Department of Electronic Engineering, Kyung Hee University, Yongin-si, Gyeonggido 17104, Korea. (email: a.nallanathan@qmul.ac.uk).}
}

\maketitle
\begin{abstract}
Diffusion model-based channel estimators have shown impressive performance but suffer from high computational complexity because they rely on iterative reverse sampling. This paper proposes a sampling-free diffusion transformer (DiT)-based channel estimator, termed SF-DiT-CE, for low-complexity MIMO channel estimation. Exploiting angular-domain sparsity of MIMO channels, we train a lightweight DiT to directly predict the true channels from their perturbed observations and noise levels. At inference, we first obtain an initial channel estimate using the least-squares (LS) method, which can be viewed as the true channel corrupted by Gaussian noise. The DiT then takes this estimate and its corresponding noise scale as inputs to recover the channel in a single forward pass, eliminating iterative sampling. Numerical results demonstrate that our method achieves superior estimation accuracy and robustness with significantly lower complexity than state-of-the-art baselines.
The code is available at: https://github.com/c-res/SF-DiT-CE
\end{abstract}

\section{Introduction}
Efficient and accurate channel estimation is essential for realizing high throughput and reliability promised by multiple-input multiple-output (MIMO) systems.
Classical estimators such as least squares (LS) and linear minimum mean square error (LMMSE) are widely adopted for their computational simplicity, but their performance often degrades in high-dimensional and non-stationary environments due to noise sensitivity or reliance on stationary channel statistics \cite{hampton2013introduction}.

To overcome limitations of classical estimators, deep learning (DL) approaches, e.g., \cite{8752012}, proposed to learn a direct mapping from received pilots to channels via supervised training. While effective, they typically require massive labeled training data that are costly to acquire in practice. Moreover, their supervised nature ties performance to the specific propagation conditions, such as signal-to-noise ratio (SNR), seen during training, limiting applicability under mismatched scenarios.

To circumvent labeled data dependency, generative adversarial network (GAN)-based estimators \cite{9316250} have been explored to learn the underlying channel prior and reconstruct channels from noisy observations. However, the training of GANs is susceptible to instability and mode collapse, which limits their ability to capture diverse channel conditions.

Recently, diffusion models have emerged as powerful generative models for learning expressive data-driven priors and have been successfully applied to MIMO channel estimation, substantially outperforming conventional DL- and GAN-based estimators \cite{9957135, 10946972, 11297769, 10705115, liu2025flow}. Existing diffusion-based channel estimation methods can be broadly categorized into variance-exploding (VE), variance-preserving (VP), and flow-based approaches. VE-based methods typically train a score model to learn the gradient of the logarithm of the MIMO channel distribution and then use it as a learned prior for inference.
For example, the estimator in \cite{9957135} performed iterative posterior sampling via annealed Langevin dynamics, while \cite{10946972} proposed a score-based variational inference scheme to accelerate estimation.
In addition, \cite{11202793} accelerated inference via step-skipping and also considered a Tweedie's formula-based single-step denoiser as an extreme case, while \cite{11016071} further reduced the reverse process to approximately 10-15 diffusion steps. In contrast, VP-based methods train a denoiser under a VP corruption process to capture the channel prior and estimate channels through iterative posterior sampling \cite{11297769}.
The authors in \cite{10705115} further reduced inference cost by initializing inference with an LS estimate and directly denoising it using a VP diffusion model to obtain the channel estimate.
More recently, flow model-based channel estimators, such as \cite{liu2025flow}, leveraged the learned velocity field as a data-driven prior for improved efficiency.

However, existing diffusion model-based channel estimators typically rely on iterative reverse sampling, requiring tens or hundreds of neural function evaluations (NFEs) for satisfactory performance. This high computational overhead and inference latency (i.e., the wall-clock runtime required for model inference to estimate the channel from the received pilot signal) hinder their deployment in real-time wireless systems.
To address these challenges, we propose a sampling-free diffusion transformer (DiT)-based channel estimator (SF-DiT-CE) for low-complexity MIMO channel estimation, which requires only a single NFE to recover MIMO channels from noisy pilot observations. Leveraging the angular-domain sparsity of MIMO channels, we train a lightweight DiT model using the VE framework to directly predict the true channels from their perturbed observations and noise levels.
This strategy reduces the learning difficulty and enhances generalization. At inference, the DiT model refines an initial LS estimate in a single forward pass (i.e., one NFE) to reconstruct the MIMO channel, eliminating iterative reverse sampling. Experimental results show that, compared to state-of-the-art channel estimators, our approach achieves up to a 4.3 dB reduction in normalized mean square error (NMSE) with significantly lower inference latency, while remaining robust to distributional shifts between training and testing environments.

\section{System Model and Preliminaries}

\subsection{MIMO Channel Estimation}
Consider a point-to-point MIMO communication system in which a transmitter equipped with $N_t$ antennas sends $N_p$ pilot symbols to a receiver with $N_r$ antennas for channel estimation. The received pilot signal is given by
\begin{align}
\mathbf{Y} = \mathbf{H}\mathbf{P}  + \mathbf{N},
\end{align}
where $\mathbf{H} \in \mathbb{C}^{N_r \times N_t}$ denotes the channel state information (CSI) matrix, $\mathbf{P} \in \mathbb{C}^{N_t \times N_p}$ is the known pilot matrix, and $\mathbf{N} \in \mathbb{C}^{N_r \times N_p}$ represents additive white Gaussian noise (AWGN) with variance $\sigma^2$.
Similar to \cite{10705115, liu2025flow}, this work considers the full-pilot setting $N_p = N_t$ and chooses $\mathbf{P}$ as a unitary discrete Fourier transform (DFT) matrix such that $\mathbf{P}\mathbf{P}^H = \mathbf{I}$.
The channel estimation task is to recover $\mathbf{H}$ from the observation $\mathbf{Y}$ given the known pilot matrix $\mathbf{P}$.

\subsection{Diffusion-Based Learning of MIMO Channel Priors}\label{subsec:diffusion_preliminary}
Let $p_X$ denote the unknown data distribution of $\mathbf{X}$, e.g., the CSI data. Diffusion models implicitly learn $p_X$ by a forward noising process that gradually perturbs clean data $\mathbf{X}_0 \sim p_X$ (with $\mathbf{X}_0 = \mathbf{X}$) into noisy latent variables $\mathbf{X}_t$ ($1\le t\le T$) via additive Gaussian noise, and a backward denoising process. Depending on the noise injection rule, diffusion models are commonly classified as VP and VE formulations \cite{song2020score}.
In VP diffusion, noise is injected while preserving the total variance over time.
The forward process is defined as
\begin{align}\label{eq:VP_corruption}
\mathbf{X}_t = \sqrt{\bar{\alpha_t}} \mathbf{X}_0 + \sqrt{1 - \bar{\alpha}_t}  \bm{\epsilon}, \bm{\epsilon} \sim \mathcal{N}(\bm{0}, \mathbf{I}),
\end{align}
where $\bar{\alpha}_t \in (0,1)$ denotes the noise-schedule variable at diffusion step $t$. As $t$ increases, $\bar{\alpha}_t$ decreases monotonically, causing the signal component to gradually diminish while the noise component becomes dominant. In VE diffusion, noise variance increases over time while the signal remains unscaled:
\begin{align}\label{eq:VE_corruption}
\mathbf{X}_t = \mathbf{X}_0 + \sigma_t \bm{\epsilon},
\end{align}
where $\sigma_t >0$ is the noise standard deviation at step $t$.

Given the forward process, diffusion models can be trained with different prediction objectives, including:
1) $\bm{\epsilon}$-prediction: The network predicts the added Gaussian noise $\bm{\epsilon}$ in $\mathbf{X}_t$ at diffusion step $t$ or noise level $\sigma_t$ \cite{song2020score, NEURIPS2022_a98846e9}.
2) $\mathbf{V}$-prediction: The network predicts flow velocity $\mathbf{V}_t = \frac{d}{d t}\mathbf{X}_t$. In the VE framework, a common velocity target is the derivative with respect to the noise scale, i.e., $\mathbf{V}_t =\frac{d}{d \sigma_t} \mathbf{X}_t = \frac{\mathbf{X}_t - \mathbf{X}_0}{\sigma_t} = \bm{\epsilon}$ \cite{lipman2022flow}.
3) $\mathbf{X}$-prediction: The network directly predicts $\mathbf{X}_0$ from noisy data $\mathbf{X}_t$ and the diffusion step $t$ or noise level $\sigma_t$ \cite{li2025back}.
Once the forward noising process is specified, these objectives are inter-convertible through deterministic algebraic relationships.

\section{Sampling-Free Diffusion Transformer for Channel Estimation}
This section presents our SF-DiT-CE, which streamlines MIMO channel estimation by requiring only a single NFE.
\subsection{Training of SF-DiT-CE Model}
As discussed in Section \ref{subsec:diffusion_preliminary}, by choosing the forward noise injection framework (VP or VE) and the network prediction objective, a diffusion model can be trained to learn a MIMO channel prior for channel estimation.
In this work, we train our SF-DiT-CE model using the following design choices:
\begin{itemize}
  \item First, we adopt the VE framework for noise injection. This choice directly aligns the diffusion forward noising process with the LS estimate, as shown in Section \ref{subsec:proposed_estimator}. Such alignment avoids the mismatch between training corruption and the inference denoising process for channel estimation. The resulting benefits are demonstrated experimentally in Section \ref{sec:simulations}.

  \item Second, we train SF-DiT-CE to directly predict the true channel, i.e., $\mathbf{X}$-prediction. This choice is motivated by the manifold assumption, which shows that high-dimensional natural data usually lie on a low-dimensional manifold \cite{li2025back}. MIMO channels share this property due to limited scattering and spatial correlation.
      Unlike the off-manifold targets used in $\bm{\epsilon}$- and $\mathbf{V}$-prediction, the true channel lies on-manifold, making it easier to learn and more training-efficient. The effectiveness of this choice is demonstrated in Section \ref{sec:simulations}.
\end{itemize}

With these design choices, we train SF-DiT-CE on an MIMO channel dataset $\mathcal{D}_H=\{\mathbf{H}_i\}_{i=1}^Q$ containing $Q$ channel realizations.
To further exploit MIMO channel sparsity and reduce learning difficulty, we transform each channel $\mathbf{H} \in \mathcal{D}_H$ into the angular domain via a spatial Fourier transform, i.e.,
\begin{align}\label{eq:spatial_angular}
\mathbf{H}_{\rm{ang}} = \mathcal{F}(\mathbf{H}) = \mathbf{U}_r^H \mathbf{H} \mathbf{U}_t,
\end{align}
where $\mathbf{U}_r$ and $\mathbf{U}_t$ denote the receive and transmit DFT matrices, respectively.
Considering neural networks operate on real-valued data, we convert each complex-valued angular-domain channel $\mathbf{H}_{\rm{ang}}$ into a two-channel 2D image $\mathbf{X}$, i.e.,
\begin{align}\label{eq:complex_to_real}
\mathbf{X} = [\mathfrak{R}(\mathbf{H}_{\rm{ang}}), \mathfrak{J}(\mathbf{H}_{\rm{ang}})] \in \mathbb{R}^{2 \times N_r \times N_t},
\end{align}
where $\mathfrak{R}(\mathbf{H}_{\rm{ang}})$ and $\mathfrak{J}(\mathbf{H}_{\rm{ang}})$ denote the real and imaginary components of $\mathbf{H}_{\rm{ang}}$, respectively.

After preprocessing, we train a diffusion model $f_{\bm{\theta}}:(\mathbf{X}_t,\sigma_t)\mapsto \widehat{\mathbf{X}}_0$ to learn the MIMO channel prior in the angular domain, where $\bm{\theta}$ denotes the parameters of the diffusion model. Here, $\mathbf{X}_t=\mathbf{X}+\sigma_t\bm{\epsilon}$ denotes the perturbed CSI image at diffusion step $t$ generated by the VE process in \eqref{eq:VE_corruption}, with $\bm{\epsilon}\sim\mathcal{N}(\mathbf{0},\mathbf{I})$, and $\widehat{\mathbf{X}}_0$ denotes the estimated CSI image.
In this work, we adopt the log-normal noise schedule \cite{NEURIPS2022_a98846e9}, i.e., $\ln (\sigma_t) \sim \mathcal{N}(\lambda_{\rm{mean}},\lambda_{\rm{std}}^2)$.
Specifically, we first sample $\ln (\sigma_t)$ from $\mathcal{N}(\lambda_{\rm{mean}},\lambda_{\rm{std}}^2)$ and then compute $\sigma_t = e^{\ln (\sigma_t)}$.
The diffusion model takes $(\mathbf{X}_t, \sigma_t)$ as input and predicts the CSI image $\widehat{\mathbf{X}}_0 = f_{\bm{\theta}}(\mathbf{X}_t, \sigma_t)$.
Then, the model parameters $\bm{\theta}$ are optimized via stochastic gradient descent on the loss function.
Note that, although we train the diffusion model $f_{\bm{\theta}}$ to predict the true channel (i.e., $\mathbf{X}$-prediction),
the loss can be defined using any of three targets: the ground-truth $\mathbf{X}_0$ ($\mathbf{X}$-loss) or the equivalent diffusion targets, namely noise ($\bm{\epsilon}$-loss) or velocity ($\mathbf{V}$-loss).
This work adopts the $\mathbf{V}$-loss. We compute the predicted velocity based on the network prediction $\widehat{\mathbf{X}}$ as $\widehat{\mathbf{V}}_t = \frac{\mathbf{X}_t - \widehat{\mathbf{X}}_0}{\sigma_t}$,
and define the loss as
\begin{align}\label{eq:loss_func}
\mathcal{L} = \left\| \widehat{\mathbf{V}}_t - \mathbf{V}_t \right\|_2^2 = \Big\| \frac{\mathbf{X}_t - \widehat{\mathbf{X}}_0}{\sigma_t} - \bm{\epsilon} \Big\|_2^2,
\end{align}
where $\mathbf{V}_t=\bm{\epsilon}$ is the ground-truth velocity under VE corruption.
The effectiveness of this loss function is validated experimentally in Section \ref{sec:simulations}. For clarity, the overall training procedure is summarized in Algorithm \ref{alg:train_DiT}.

\begin{algorithm}[t]
\caption{Training of SF-DiT-CE}
\label{alg:train_DiT}
\begin{algorithmic}[1]
\State \textbf{Inputs:}  Training dataset $\mathcal{D}_{H}=\left\{\mathbf{H}_i \right\}_{i = 1}^Q$, batch size $B$.
\State  Convert $\mathbf{H}_i \in \mathcal{D}_H$ to angular domain using \eqref{eq:spatial_angular}, reshape into a 2D CSI image using \eqref{eq:complex_to_real}, and form $\mathcal{D}_X=\left\{\mathbf{X}_i \right\}_{i = 1}^Q$.
\Repeat
    \State Randomly sample a batch of $B$ data from $\mathcal{D}_X$.
    \State \parbox[t]{\dimexpr\linewidth-\algorithmicindent}{For each $\mathbf{X}_i$ in the batch, randomly sample $\sigma_t$ with $\ln (\sigma_t) \sim \mathcal{N}(\lambda_{\rm{mean}},\lambda_{\rm{std}}^2)$ and $\bm{\epsilon} \sim \mathcal{N}(0,  \mathbf{I})$ to perturb $\mathbf{X}_i$ based on \eqref{eq:VE_corruption}.}
    \State Estimate the channel as $\widehat{\mathbf{X}}_0 = f_{\bm{\theta}}(\mathbf{X}_t, \sigma_t)$
    \State Compute the training loss according to \eqref{eq:loss_func}.
    \State \parbox[t]{\dimexpr\linewidth-\algorithmicindent}{Take gradient descent on the loss to update the model.}
\Until{the diffusion transformer model $f_{\bm{\theta}}$ converges.}
\end{algorithmic}
\end{algorithm}

\begin{figure}[ht]
\centering
\includegraphics[width=0.36\textwidth]{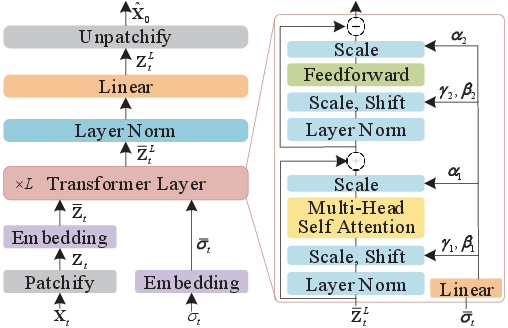}\\
\caption{The architecture of the adopted DiT model.}
\label{fig:DiT_architecture}
\end{figure}

\subsection{Network Architecture}
To reduce training and inference complexity, this work adopts a lightweight DiT \cite{Peebles_2023_ICCV} to learn the MIMO channel prior, as illustrated in Fig. \ref{fig:DiT_architecture}.
Given the input pair $(\mathbf{X}_t, \sigma_t)$, the model first transforms $\mathbf{X}_t \in \mathbb{R}^{2 \times N_r \times N_t}$ into $M=\frac{N_r}{\tau}\times\frac{N_t}{\tau}$ 2D patches and projects each patch into a $d$-dimensional token, yielding $\mathbf{Z}_t \in \mathbb{R}^{M \times d}$, where $\tau$ denotes the patch size.
A 2D sinusoidal positional embedding is then added to obtain the embedded token sequence $\bar{\mathbf{Z}}_t$.
In parallel, the noise level $\sigma_t \in \mathbb{R}$ is mapped through a sinusoidal embedding and a linear layer to produce adaptive modulation parameters for each transformer block, including scale-shift pairs for layer normalization and gating factors for the residual branches.
The embedded tokens are then processed by a stack of $L$ transformer blocks.
Finally, the output tokens are normalized, linearly projected back to patch pixels, and unpatchified to reconstruct the estimated CSI image $\widehat{\mathbf{X}}_0 \in \mathbb{R}^{2 \times N_r \times N_t}$.

\subsection{Inference of SF-DiT-CE for Channel Estimation}\label{subsec:proposed_estimator}
To address the high computational complexity of diffusion-based channel estimators caused by iterative reverse sampling, we propose a sampling-free framework that exploits the learned diffusion prior through a single forward pass of the DiT model.
Firstly, we  obtain a LS estimate of the channel from the received pilot signal $\mathbf{Y}$ as:
\begin{align}\label{eq:LS_solution}
\widehat{\mathbf{H}}_{\rm{LS}} = \mathbf{Y} \mathbf{P}^H = \mathbf{H} + \mathbf{N}\mathbf{P}^H.
\end{align}
%where $\mathbf{N}\mathbf{P}^H$ is AWGN with variance $\sigma^2$, due to the unitary property of the pilot matrix $\mathbf{P}$.
Since our DiT model is trained on angular-domain channel data, we transform $\widehat{\mathbf{H}}_{\rm{LS}}$ to the angular domain as
\begin{align}\label{eq:LS_angular}
\widehat{\mathbf{H}}_{\rm{LS, ang}} = \mathcal{F}(\widehat{\mathbf{H}}_{\rm{LS}}) & = \mathbf{U}_r^H \mathbf{H} \mathbf{U}_t + \mathbf{U}_r^H \mathbf{N}\mathbf{P}^H \mathbf{U}_t  \notag \\
& = \mathbf{H}_{\rm{ang}} + \widetilde{\mathbf{N}},
\end{align}
where the noise term $\widetilde{\mathbf{N}}  = \mathbf{U}_r^H \mathbf{N}\mathbf{P}^H \mathbf{U}_t$. Because $\mathbf{P}$, $\mathbf{U}_r$, and $\mathbf{U}_t$ are unitary, $\widetilde{\mathbf{N}}$ is a unitary rotation of $\mathbf{N}$. Moreover, since AWGN is invariant under unitary transformations, $\widetilde{\mathbf{N}}$ remains AWGN with the same variance $\sigma^2$.

Hence, we transform $\widehat{\mathbf{H}}_{\rm{LS, ang}}$ to a 2D image $\widehat{\mathbf{X}}_{\rm{LS, ang}} = [\mathfrak{R}(\widehat{\mathbf{H}}_{\rm{LS, ang}}), \mathfrak{J}(\widehat{\mathbf{H}}_{\rm{LS, ang}})]$.
After that, we feed $\widehat{\mathbf{X}}_{\rm{LS, ang}}$ and the corresponding noise level $\sigma$ into the DiT model to predict the true channel in a single forward pass, i.e.,
\begin{align}
\widehat{\mathbf{X}}_{\rm{ang}} = f_{\bm{\theta}}(\widehat{\mathbf{X}}_{\rm{LS, ang}}, \sigma).
\end{align}
Following that, we convert $\widehat{\mathbf{X}}_{\rm{ang}}$ into the complex domain, i.e., $\widehat{\mathbf{H}}_{\rm{ang}} = \widehat{\mathbf{X}}_{\rm{ang}}[0,:,:] + j \times \widehat{\mathbf{X}}_{\rm{ang}}[1,:,:]$.
Finally, $\widehat{\mathbf{H}}_{\rm{ang}}$ is transformed back to the spatial domain as
\begin{align}
\widehat{\mathbf{H}} = \mathcal{F}^{-1}(\widehat{\mathbf{H}}_{\rm{ang}}) = \mathbf{U}_r \widehat{\mathbf{H}}_{\rm{ang}} \mathbf{U}_t^H,
\end{align}
which is the estimated channel.

\begin{algorithm}[t]
\caption{SF-DiT-CE for MIMO Channel Estimation}
\label{alg:ch_estimation}
\begin{algorithmic}[1]
\State \textbf{Inputs:} $\mathbf{Y}$, $\mathbf{P}$, $f_{\bm{\theta}}$, $\sigma^2$
\State Compute the LS estimate $\widehat{\mathbf{H}} = \mathbf{Y} \mathbf{P}^H$ and convert it to angular domain as $\widehat{\mathbf{H}}_{\rm{LS, ang}} = \mathcal{F}(\widehat{\mathbf{H}}_{\rm{LS}})$
\State Estimate the channel as $\widehat{\mathbf{H}}_{\rm{ang}} = f_{\bm{\theta}}(\widehat{\mathbf{H}}_{\rm{LS, ang}}, \sigma)$
\State Transform back to spatial domain as $\widehat{\mathbf{H}} = \mathcal{F}^{-1}(\widehat{\mathbf{H}}_{\rm{ang}})$
\State \textbf{Output:} Estimated CSI $\widehat{\mathbf{H}}$.
\end{algorithmic}
\end{algorithm}

For clarity, Algorithm \ref{alg:ch_estimation} summarizes the proposed channel estimation procedure.
The following remark highlights the advantages of our design choices.

\begin{rem}\label{remark:remark1}
According to \eqref{eq:LS_angular}, the angular-domain LS estimate $\widehat{\mathbf{H}}_{\rm LS,ang}$ has the additive form of the true channel corrupted by AWGN at noise level $\sigma$, which is exactly consistent with the VE forward process in \eqref{eq:VE_corruption}. In contrast, the VP-based method in \cite{10705115} aligns the LS estimate to the diffusion process by normalizing the LS estimate and selecting the reverse starting step through matching the observation's SNR to the diffusion model's SNR.
Since the practical observation's SNR is random and continuous-valued, whereas the VP diffusion model is defined on a discrete set of SNR levels, the observation SNR may not exactly match any diffusion step.
As a result, the reverse process in \cite{10705115} may introduce residual misalignment between the LS estimate and the VP corruption process. This mismatch can bias denoising performance.
Our VE formulation avoids this mismatch. Moreover, since the proposed SF-DiT-CE directly predicts the true channel from the noisy observation, it can accurately recover the channel in a single forward pass.
\end{rem}

\section{Numerical Results}\label{sec:simulations}
This section evaluates the proposed SF-DiT-CE. Estimation accuracy is measured by NMSE, i.e., $\rm{NMSE}[dB] = \mathbb{E}_{\mathbf{H}} \Big[ 10\log_{10}\frac{ \big\| \mathbf{H} - \widehat{\mathbf{H}} \big\|_F^2} {\left\| \mathbf{H} \right\|_F^2}\Big]$, where $\mathbf{H}$ and $\widehat{\mathbf{H}}$ denote the true and estimated channel matrices, respectively.
The DiT model uses $L=2$ Transformer layers with the hidden dimension $d=128$, patch size $\tau=4$, and 8 attention heads.
The noise schedule parameters are set to $\lambda_{\rm{mean}}=-1.2$ and $\lambda_{\rm{std}}=1.2$.
We generate training and test data using the clustered delay line (CDL) models in the MATLAB 5G Toolbox, following the 3GPP TR 38.901 specification \cite{3gpp2020study}. Specifically, we create 10000 channel realizations for each of the CDL-C and CDL-D profiles for training, and 1000 realizations per profile for testing.
Both the transmitter and receiver employ uniform linear arrays with $(N_r,N_t)=(64,16)$, with other settings following \cite{10946972}.
SF-DiT-CE is trained for 1000 epochs on each training dataset.
All experiments are conducted on a Linux server equipped with an NVIDIA RTX 4500 GPU and an Intel Xeon Gold 5418Y CPU.

We compare the proposed method with the following baselines:
1) LS: The solution is given in \eqref{eq:LS_solution}.
2) LMMSE \cite{hampton2013introduction}: It assumes that the vectorized channel follows a complex Gaussian distribution, i.e., $\mathbf{h}=\mathrm{vec}(\mathbf{H}) \sim \mathcal{CN}(\bm{\mu}, \mathbf{C})$, with $\bm{\mu}$ and $\mathbf{C}$ estimated from the training dataset. The LMMSE estimate is then given by $\widehat{\mathbf{H}}=\mathrm{unvec}(\widehat{\mathbf{h}})$, where $\widehat{\mathbf{h}}=\bm{\mu}+\mathbf{C}\big(\mathbf{C}+\sigma^2\mathbf{I}\big)^{-1}\big(\mathrm{vec}(\mathbf{Y}\mathbf{P}^H)-\bm{\mu}\big)$. Here $\mathrm{vec}(\cdot)$ and $\mathrm{unvec}(\cdot)$ denote vectorization and de-vectorization, respectively.
3) DMCE \cite{10705115}: It trains a CNN-based estimator using VP framework and performs iterative denoising for channel estimation.
4) DMCE (with DiT): A variant of \cite{10705115} that replaces its CNN with the DiT used in this work, while keeping other settings unchanged.
5) Score model-based Langevin sampling approach (Score) \cite{9957135}: It trains a RefineNet to learn the channel score function and performs Langevin dynamics for channel estimation.
6) Our approach with VP perturbation:
This variant follows the proposed method but perturbs channels using the VP framework during training, and conditions the model on the LS estimate and SNR at inference.
7) Spatial-domain variant of our approach: This variant operates directly on spatial-domain channel data without angular-domain transforms.
%The implementations of DMCE and the score-based method are reproduced strictly following their publicly available codebases.

\begin{figure}[ht]
\centering
\includegraphics[width=0.41\textwidth]{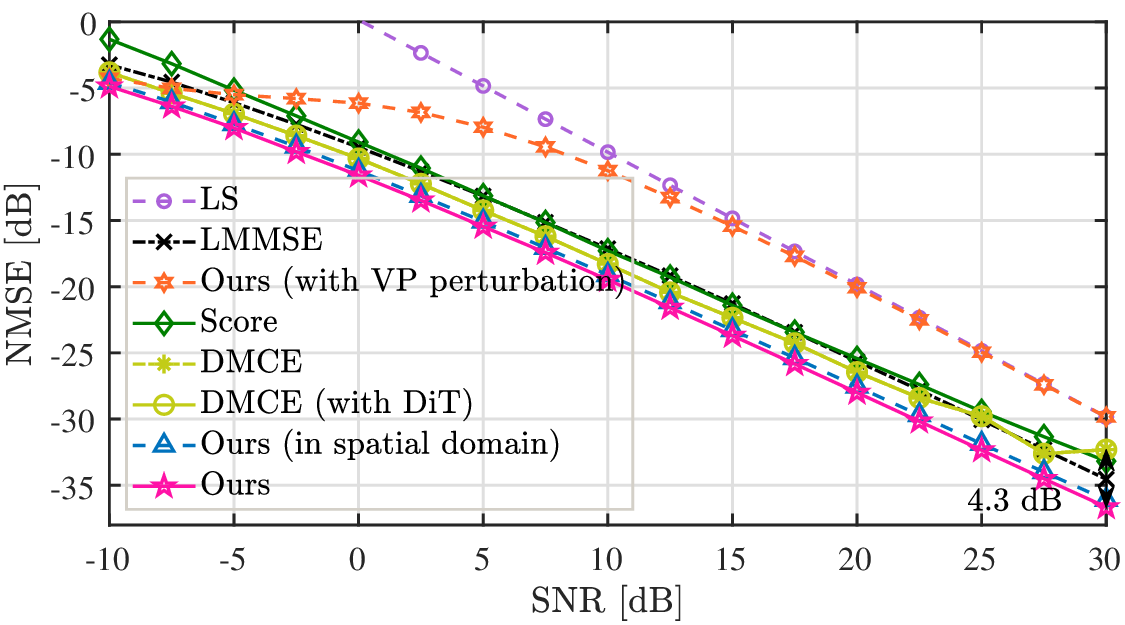}\\
\caption{Comparison of different channel estimators on CDL-C dataset.}
\label{fig:snr_nmse_algs}
\end{figure}

Fig. \ref{fig:snr_nmse_algs} reports the NMSE versus SNR for different channel estimators on the CDL-C test dataset, where all diffusion-based estimators are trained on the CDL-C training set.
The proposed method achieves the lowest NMSE across the entire SNR range. Compared with the best baseline, i.e., DMCE, it reduces the NMSE by up to 4.3 dB.
This gain mainly comes from the VE corruption and direct true-channel prediction, which align the initial LS estimate with the diffusion forward process, avoid the input-process mismatch discussed in Remark \ref{remark:remark1}, and enable sampling-free inference.
In addition, DMCE and DMCE (with DiT) exhibit similar performance, indicating that the gain of the proposed method does not mainly arise from the larger model size.
%By contrast, DMCE adopts a VP formulation and matches the LS estimate to the diffusion process through discrete SNR-step selection, which may introduce residual mismatch.
Note that, DMCE shows fluctuations at high SNR, likely due to its discrete linear noise schedule, which is sensitive to both schedule hyperparameters and channel noise. In comparison, our method uses a log-continuous noise schedule, making it more robust to noise-level variations.
Moreover, the proposed approach significantly outperforms its VP-perturbation counterpart, further highlighting the importance of consistency between the training corruption and the inference denoising processes.

\begin{table}[ht]
\footnotesize
\centering
\caption{Complexity and runtime comparison}
\label{tab:complexity_compare}
\begin{tabular}{|>{\centering\arraybackslash}m{0.8cm}|>{\centering\arraybackslash}m{1.23cm}|>{\centering\arraybackslash}m{0.8cm}|>{\centering\arraybackslash}m{3.1cm}|>{\centering\arraybackslash}m{0.5cm}|}
\hline
Method & \# Params. & SNR & Runtime [ms] (CPU/GPU) & NFE\\ \hline
LS & - & All & 0.03 / 0.02 & -  \\ \hline
LMMSE & - & All & 8.79 / 2.19 & -  \\ \hline
\multirow{5}{*}{DMCE} & \multirow{5}{*}{$5.5 \times 10^4$}
& -10 dB & 68.6 / 27.7 & 58  \\ \cline{3-5}
 &  & 0 dB & 36.2 / 14.8 & 28 \\ \cline{3-5}
 &  & 10 dB & 17.6 / 6.25 & 9 \\ \cline{3-5}
 &  & 20 dB & 5.83 / 2.72 & 3 \\ \cline{3-5}
 &  & 30 dB & 2.47 / 1.14 & 1 \\ \hline
Score & $5.89\times 10^6$ & All & $1.1\times 10^5$ / $5.7 \times 10^4$ & 6933 \\ \hline
Ours & $0.67\times 10^6$ & All & 8.41 / 1.83 & 1 \\ \hline
\end{tabular}
\end{table}

Table \ref{tab:complexity_compare} compares the complexity of classical and diffusion-based channel estimators. LS and LMMSE have the lowest latency, but they typically deliver inferior estimation accuracy relative to diffusion-based methods, as shown in Fig. \ref{fig:snr_nmse_algs}.
For diffusion-based estimators, the runtime is mainly determined by the NFE and the complexity of the denoising network.
DMCE employs a lightweight CNN, yet its iterative refinement yields SNR-dependent NFE, decreasing from 58 at -10 dB to 1 at 30 dB, with CPU latency varying from 68.6 ms to 2.47 ms.
The score-based method is the most computationally intensive, requiring 6933 NFEs with a large RefineNet backbone.
In contrast, the proposed method is sampling-free and requires only one single forward pass, leading to a constant NFE of 1 and SNR-independent latency of 8.41 ms on CPU and 1.83 ms on GPU.
Although its network has more parameters than the CNN in DMCE, it is significantly faster in the low-SNR regime and provides a practical diffusion-based estimator with substantially reduced complexity and improved performance compared with iterative samplers.

\begin{figure}[ht]
\centering
\includegraphics[width=0.41\textwidth]{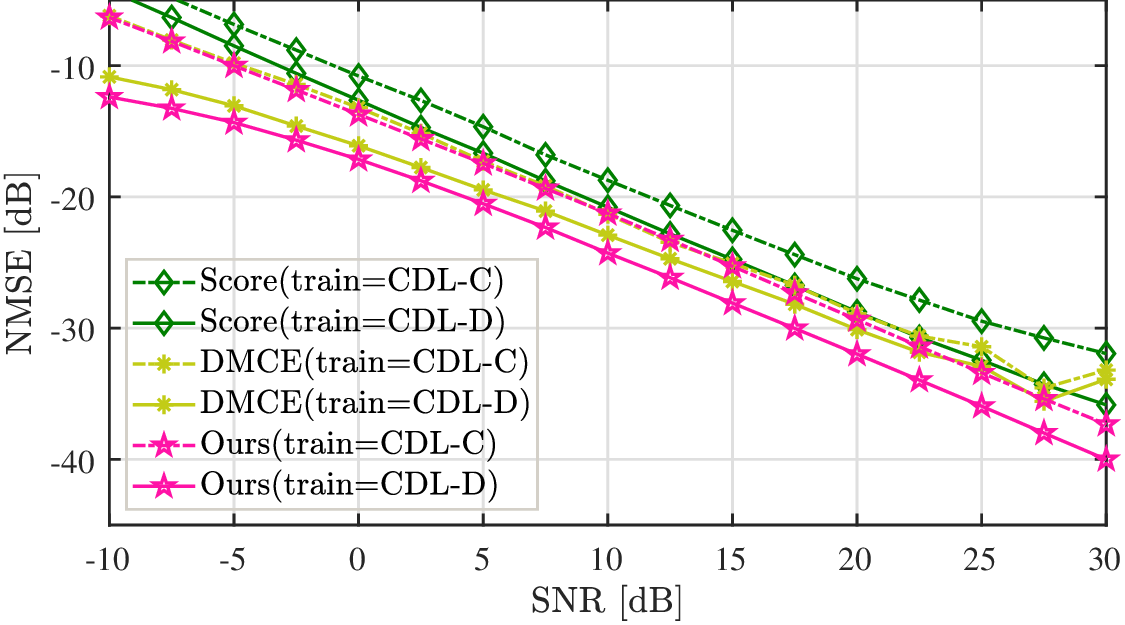}\\
\caption{Comparison of diffusion-based channel estimators on CDL-D dataset.}
\label{fig:snr_algs_train_test_shift}
\end{figure}

Fig. \ref{fig:snr_algs_train_test_shift} evaluates the robustness of diffusion-based estimators against distributional shifts by training them on the CDL-C dataset and testing them on CDL-D.
As a reference, we include in-distribution results where each estimator is trained and tested on CDL-D.
The proposed method consistently achieves the lowest NMSE, outperforming all baselines across the entire SNR range.
Remarkably, the proposed method under distributional shift (trained on CDL-C) still outperforms the Score and DMCE baselines even when they operate without distribution shift (trained on CDL-D).
While the proposed approach trained on CDL-C dataset is surpassed by the in-distribution DMCE at lower SNRs ($-10$ to $20$ dB), it is important to note that our model operates under a distributional mismatch while DMCE does not.
Furthermore, our method maintains a significantly lower inference latency in this low SNR region. As shown in Table \ref{tab:complexity_compare}, our approach achieves an acceleration of approximately 27.8$\times$ over DMCE at an SNR of $-10$ dB on the CPU and 24.3$\times$ on the GPU.

\begin{figure}[t]
\centering
\subfigure[]{\includegraphics[width=0.48\linewidth]{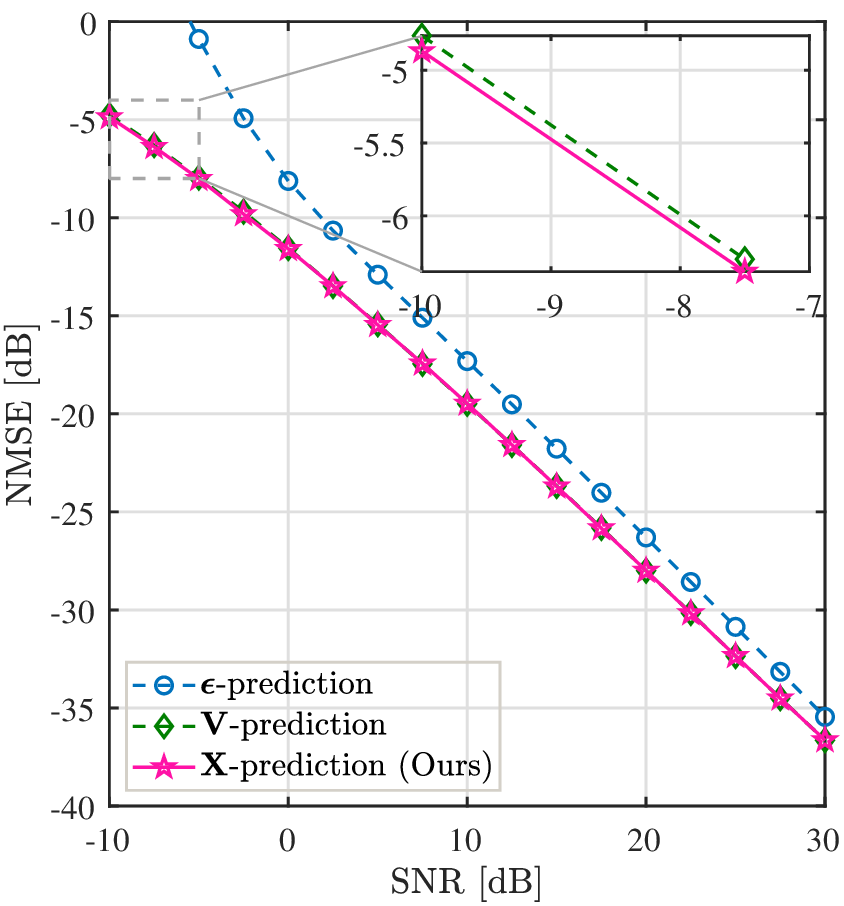}\label{fig:ablation_obj}}
\subfigure[]{\includegraphics[width=0.48\linewidth]{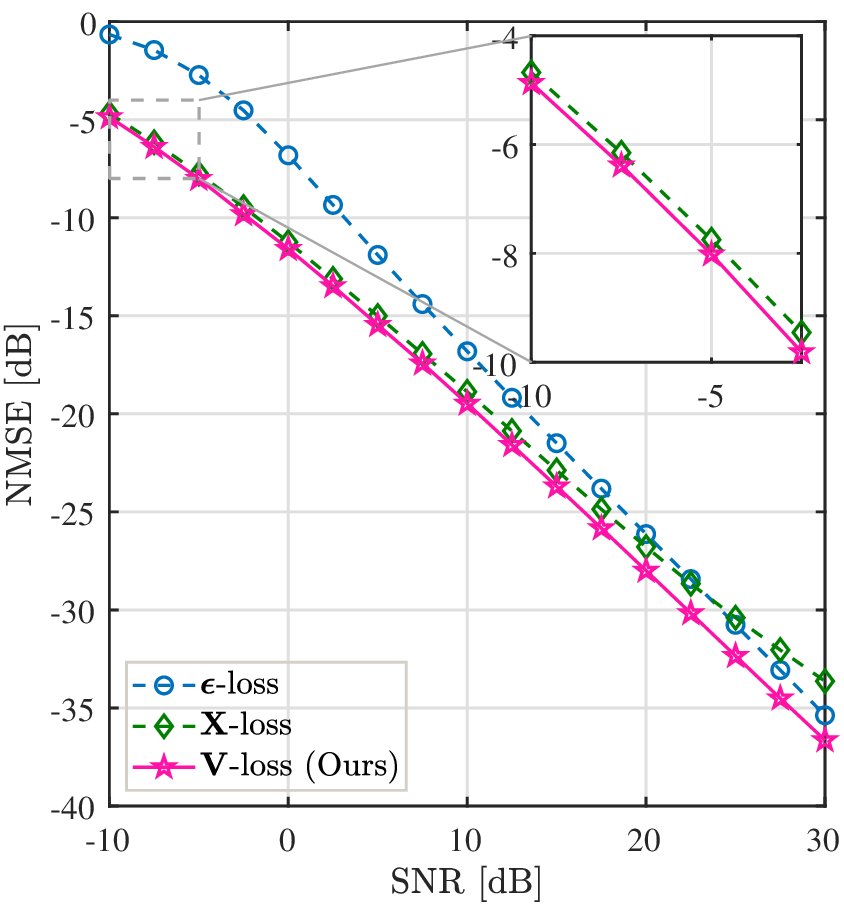}\label{fig:ablation_loss}}
\caption{Impact of (a) prediction objective and (b) loss function on NMSE of the proposed approach.}
\label{fig:ablation}
\end{figure}

Fig. \ref{fig:ablation} investigates how the network prediction objective and loss function affect the proposed channel estimator.
As shown in Fig. \ref{fig:ablation_obj}, $\mathbf{X}$-prediction consistently achieves lower NMSE than $\bm{\epsilon}$-prediction and $\mathbf{V}$-prediction.
Here, for $\bm{\epsilon}$-prediction, the channel estimate is derived via Tweedie's formula as $\widehat{\mathbf{X}}_{\rm{ang}} = \widehat{\mathbf{X}}_{\rm{LS, ang}} + \sigma^2 f_{\bm{\theta}}(\widehat{\mathbf{X}}_{\rm{LS, ang}}, \sigma)$ \cite{11202793}, while for $\mathbf{V}$-prediction, it is given by $\widehat{\mathbf{X}}_{\rm{ang}} = \widehat{\mathbf{X}}_{\rm{LS, ang}} - \sigma f_{\bm{\theta}}(\widehat{\mathbf{X}}_{\rm{LS, ang}}, \sigma)$.
This can be attributed to the fact that high-dimensional MIMO channels typically lie on a low-dimensional manifold due to limited scattering, whereas the noise and velocity targets are more off-manifold.
As a result, directly predicting the true channels reduces the required model capacity and eases learning.
Fig. \ref{fig:ablation_loss} demonstrates that $\mathbf{V}$-loss yields superior estimation accuracy compared to $\bm{\epsilon}$-loss and $\mathbf{X}$-loss.
The proposed SF-DiT-CE leverages the direct reconstruction capability of $\mathbf{X}$-prediction alongside the stable gradient flow provided by $\mathbf{V}$-loss to maximize estimation performance.

\section{Conclusion}
This work proposed a novel sampling-free DiT for low-complexity MIMO channel estimation, termed SF-DiT-CE. It achieves accurate channel estimation with only a single NFE, thereby eliminating the iterative reverse sampling inherent in existing diffusion-based estimators and substantially reducing inference latency. Experimental results demonstrate that SF-DiT-CE consistently outperforms state-of-the-art methods in estimation accuracy and exhibits strong robustness under train-test distribution shifts. Establishing rigorous theoretical guarantees for the proposed framework is left for future work.

\bibliographystyle{IEEEtran}
\bibliography{IEEEabrv,cited}

\end{document}